\documentstyle[preprint,aps,epsfig]{revtex}

\def\be{\begin{equation}}
\def\ee{\end{equation}}

\begin{document}
\draft
\title{Heterogeneous aggregation in binary colloidal alloys} 
\author{Andrew C. T. Wong and K. W. Yu}
\address{Department of Physics, The Chinese University of Hong Kong, \\
         Shatin, New Territories, Hong Kong, China }
 
\maketitle

\begin{abstract}
Molecular dynamics (MD) simulation has been employed to study the 
nonequilibrium structure formation of two types of particles 
in a colloidal suspension, driven by type-dependent forces. 
We examined the time evolution of structure formation as well as on 
the structural properties of the 
resulting aggregation by studying the radial distribution function (RDF).
The resulting aggregation is well described by a binary colloidal gelation.
We compared the structural properties to those for one type of particles.
From the MD results, it is evident that there are significant differences 
between the RDF's of the two cases. 
Moreover, we found that the average coordination number is generally larger 
in the monodisperse case for all area fractions considered. Thus, by means
of heterogeneous aggregation, it is possible to obtain a wide variety of
structures while more close-packed structures are formed for monodisperse 
colloidal aggregation.
\end{abstract} 
\vskip 5mm

\section{Introduction}

Soft condensed matter are materials which can easily be deformed by 
external stresses, electric and magnetic fields, 
or even by thermal fluctuations \cite{Chaikin}. 
These materials typically possess structures which are much larger than 
atomic or molecular scales; the structure and dynamics at the mesoscopic 
scales determine the macroscopic physical properties. 
The goal of our research is to model and understand this relationship. 
These materials can be synthesized by means of colloidal self-assembly.
In what follows, we will describe a model colloidal system in which the
particles interact through type-dependent forces, so as to study 
the relationship between the interaction at the mesoscopic scales and 
the macroscopic properties.

There is a long-standing yet fundamental question in physics: 
Given all the interactions between the particles, what will be the 
resulting structure that the particles will form?
For monodisperse case, usually close-packed structures will be formed, 
e.g., hexagonal structure in two dimensions and face-centered cubic or 
hexagonal close-packed structures in three dimensions.
By allowing two different types of particles, driven by two types of forces,
we are prepared to show that a rich variety of possible structures will
be formed, as we will study in more detail below.

The plan of the paper is as follows.
In the next section, we will discuss the molecular dynamics simulation 
method. In section III, we will present the results.
Discussion and conclusion on our results will be given.

\section{Molecular Dynamics Simulation}

The colloidal system studied in this work is a two-dimensional one
consisting of circular particles of the same diameter $d$, 
suspended in a viscous fluid. There is a pairwise interparticle force 
${\bf F}_{ij}=-\nabla U({\bf r}_{ij})$ between particles $i$ and $j$,
where $U$ is a potential and ${\bf r}_{ij}={\bf r}_i - {\bf r}_j$.
The equation of motion of a particle is given by
\be
m \frac{d^2{\bf r}_{i}}{dt^2}=  \sum_{j\neq i}{\bf F}_{ij} 
 - w\frac{d {\bf r}_{i} }{dt} ,
\ee
where the first term on the right-hand side describes the total force 
acting on the particle $i$,
while the second term denotes the viscous drag (with coefficient $w$), 
exerted by the fluid.  
We have neglected the Brownian motion, which is a valid assumption for 
mesoscale objects.
The variables can be rescaled as 
$t=t_0 t^*$, ${\bf r}=d{\bf r^*}$, and ${\bf F}=F_0{\bf F^*}$,
with $t_0=m/w$ and $F_0$ being the typical magnitude of the 
interparticle force. The rescaled equation of motion can be written as
\be
{\bf \ddot{r}}_i^*=A \sum_{j\neq i}{\bf F}_{ij}^* - {\bf \dot{r}}^*_i ,
\label{eom}
\ee
where $A=F_0 t_0/w$. 
In a highly viscous medium, $A$ is very small and the particles are 
in overdamped motion, i.e., ${\bf \ddot{r}}_i^* \approx 0$. 
In this case, Eq.~(\ref{eom}) reduces to
\be
{\bf \dot{r}}^*_i=A \sum_{j\neq i}{\bf F}_{ij}^* ,
\ee
We further define $\tilde{t}^*=A t^*$ and obtain the reduced equation 
of motion for our simulation: 
\be
\frac{d {\bf r}_i^* }{d \tilde{t}^*}=\sum_{j\neq i}{\bf F}_{ij}^* .
\ee
Thus we have a set of simultaneous first-order differential equations 
to solve. In the overdamped situation, the final structure is independent 
of $A$, but the time for the structure formation is inversely proportional 
to $A$.
The initial configurations consist of $N$ particles randomly dispersed in
a square simulation cell, with periodic boundary conditions being imposed
in two directions.
Thus, we include the forces acting on a particle due to all other particles
in the simulation cell as well as their periodic images, calculated within 
a square centered at that particle which is of the same size of the 
simulation cell. The simulations are run for a sufficiently long time so
that the particles have no further movements as time goes on.

The simulation cell is set to the scales $25d \times 25d$ and we ran 
simulations of area fraction of particles  
$\Phi=0.13$, $0.19$, $0.25$, $0.31$, $0.37$ and $0.44$ 
corresponding to the number of particles 
$N=100$, 150, 200, 250, 300, 350 respectively. 
The simple Euler algorithm has been used throughout the simulation with
a time step $\delta \tilde{t}^* = 5\times 10^{-4}$. 
Larger time steps may result in unphysically large velocities because 
of serious overlapping between the particles.

We consider two cases. The first case consists of one type of particles,
interact through attractive force, while the second case consists of 
equal number of different types of particles A and B. 
The initial positions and velocities of the particles are imposed by a 
random distribution inside the simulation cell without particle overlapping.  
It is difficult to generate for $N$ larger than 350 because of serious 
particle overlapping.

 \subsection{One type of particles with attractive force}

In this case all particles attract each other at short distance and 
repel each other at long distance. The potential between the partilces is
\be
U(r)= u_0 \exp(-(r-d)/\xi_0) - u_1 \exp(-(r-d)/\xi_1).
\ee
The first term denotes a repulsion, with a range $\xi_0$ and magnitude 
$u_0$. The second term denotes an attraction, with a range $\xi_1$ and 
magnitude $u_1$. We set $\xi_0>\xi_1$ and $u_1>u_0$ so that the particles 
attract each other at short distance.  
In order to avoid too much overlap when the particles approach and finally
touch, we add a steep repulsive potential of the form $b/r^{11}$
to the interparticle potential. 
Thus, the force is calculated by the negative gradient of the potential
\be
F(r)=\frac{11b}{r^{12}}+\frac{u'}{\xi_0}\exp(-(r-d)/\xi_0)
 -\frac{u_1}{\xi_1}\exp(-(r-d)/\xi_1)
\ee
where 
$$
u'=u_0-{b\over d^{11}},\ {\rm and}\ \ 
b=\frac{u_0/\xi_0-u_1/\xi_1}{1/\xi_0 d^{11}-11/d^{12}}.
$$ 
In this way, the two particles attract each other when they approach, 
while the force vanish when $r=d$, and repel each other when they 
overlap. If we rescale the force by $b=F_0 d^{12} b^*$, $u'=F_0 d u'^*$, 
$u_0=F_0 d u_0^*$ and $u_1=F_0 d u_1'^*$, we obtain the dimensionless force:
\be
F^*(r)=\frac{11b^*}{ {r^*}^{12} }
 +\frac{u'^*}{\xi_0^*}\exp(-(r^*-1)/\xi_0^*)
 -\frac{u_1^*}{\xi_1^*}\exp(-(r^*-1)/\xi_1^*) ,
\ee
where 
$$
b^*=\frac{u_0^*/\xi_0^*-u_1^*/\xi_1^*}{1/\xi_0^* -11}\ {\rm and}\ \ 
u'^*=u_0^*-b^* .
$$  
Here $\xi_0^*$, $\xi_1^*$, $u_0^*$ and  $u_1^*$ are adjustable parameters.
In this simulation, the parameters used are $\xi_0^*=0.5$, $\xi_1^*=0.3$, 
$u_0^*=5$, and $u_1^*=6$.

 \subsection{Two type of particles}

The system consist equal number of A and B particles of total area 
fraction $\Phi$. For unlike particles, there are short range attractive 
and repulsive forces between the particles, same as for one type of 
particles above. The potential between two particles is 
\be
U_-(r)=u_0 \exp(-(r-d)/\xi_0) - u_1 \exp(-(r-d)/\xi_1),
\ee
while for the same type of particles, there is always a repulsive 
potential between them
\be
U_+(r)=u_0 \exp(-(r-d)/\xi_0).
\ee
Again, we impose a steep repulsive potential of the form $b/r^{11}$
to the interparticle potential. 
Likewise, the rescaled dimensionless force are: 
\be
F_-^*(r)=\frac{11b^*}{ {r^*}^{12} }
 +\frac{u'^*}{\xi_0^*}\exp(-(r^*-1)/\xi_0^*)
 -\frac{u_1^*}{\xi_1^*}\exp(-(r^*-1)/\xi_1^*)
\ee
\be
F_+^*(r)=\frac{11b^*}{ {r^*}^{12} }
+\frac{u'^*}{\xi_0^*}\exp(-(r^*-1)/\xi_0^*) 
\ee
respectively. The same parameters are used as in the monodisperse case.

\section{Results}

We calculate the radial distribution function to study the stucture 
of the aggregations. The normalized two dimensional 
radial distribution function is defined as \cite{Allen}
\be  
g_0(r^*)=\frac{1}{8 \Phi N \Delta }\sum_i\sum_{j\neq i} 
\frac{\delta(r^*-r_{ij}^*)}{r_{ij}^*} 
\ee
where $\Delta$ is the width of histogram of $g_0$.  $g_0$ is normalised 
such that it tends to 1 at long distances.

For the analysis of simulations of two types of particles, we introduce 
partial radial distribution functions, namely, 
$g_{AA}$ is the contribution of $g_0$ by radial distances between 
particles of type A, $g_{BB}$ the 
contributions of type B, and $g_{AB}$ the contributions of radial 
distances between different types A and B. 
It is clear that $g_0=g_{AA}+g_{BB}+g_{AB}$. The distances are counted 
over the particles in the cells and 
their periodic images, in the same way as calculating the forces.  
The calculations are averaged over 20 
ensembles for better statistics.

We also calculated the mean coordination number, counted over the 
particles and their periodic images and averaged over 20 ensembles.

1. Nonequilibrium structure formation for two types of particles, driven 
by type-dependent forces: We will focus on the time evolution of 
structure formation as well as on the structural properties of the 
resulting aggregation by studying the radial distribution function (RDF).
We plot the time series of $g_0$ of aggregation in Fig.1 with four 
panels (a)--(d), from initial to final time, both for one type and two 
types of particles (shown on the same figure). The same figure will be 
used in describing results in item 2 below.

We plot the partial RDF's ($g_{AA}$, $g_{BB}$, $g_{AB}$) as well as the total 
RDF ($g_0$) in Fig.2 with six panels (a)--(f), corresponding to the six area 
fractions studied, for the binary colloidal case.
The resulting aggregation is well described by a binary alloy gelation.

2. We will compare the structural properties to those for one type of 
particles (monodisperse case). 
We plot the RDF ($g_0$) in Fig.3 with six panels (a)--(f), 
corresponding to the six area fractions studied, for the monodisperse case.
From the results, it is evident that there are significant differences 
between the RDF's of the two cases. In this regard, we will also study 
the statistical geometry by computing the average coordination number of 
the resulting aggregation. 

We plot the average coordination number $\langle Z\rangle$ versus area 
fraction in Fig.4 both for one type and two types of particles.
We find that the average coordination number is generally larger in the 
monodisperse case for all area fractions considered. It turns out that more 
close-packed structures (in fact hexagonal) are formed for monodisperse 
colloidal aggregation.

3. We will also study the partial RDF's in more detail. By symmetry, 
the resulting binary colloidal alloy is statistically bipartite, 
i.e., $g_{AA}=g_{BB}$. That means that the resulting lattice can be 
decomposed into two statistically identical lattice. 
According to our results in Fig.2, the resulting aggregation can be 
decomposed into local square order (tetravalent sites) and local 
hexagonal order (trivalent sites). In the limit of a large attractive 
force between the different types of particles, however, 
we may enhance tetravalent sites at the expense of trivalent sites. 
In this regard, we will do additional simulations for the largest area 
fraction (corresponding to $N=350$), with a larger $u_1=12$, 
other parameters being fixed, so as to suppress the trivalent sites. 
In Fig.5, we plot the partial RDF's for the $u_1=12$ case. 
The $u_1=6$ case is plotted on the same figure for comparison.

\section{Discussion and conclusion}

Here a few comments on our results are in order.
The self-assembly of two types of particles can be realized by a recent 
experimental demonstration of DNA-assisted self-assembly of nanoparticles 
\cite{DNA}. While the present investigation has been on isotropic 
interparticle forces, it is instructive to extend the present work to 
polydisperse electrorheological (ER) fluids, in which the suspended 
particles can have different dielectric permittivities \cite{Yu}.
In ER fluids, the polarized particles aggregate under the influence of
anisotropic dipolar forces. By tuning the strength of the applied field,
it is possible to realize a structure transformation from the body-centered
tetragonal to the face-centered cubic structures \cite{Lo}. Thus, by 
considering two different types of particles driven both by isotropic and
anisotropic interparticle forces, we may obtain a diversity of structures 
with potential applications in photonic band-gap materials.

In conclusion, we have performed a detailed molecular dynamics simulation 
to study the nonequilibrium structure formation for two types of particles 
in a colloidal suspension, driven by type-dependent forces. 
We examined the time evolution of structure formation as well as on 
the structural properties of the resulting aggregation by studying the 
radial distribution function (RDF).
The resulting aggregation is well described by a binary colloidal gelation.
We find that the average coordination number is generally larger in the 
monodisperse case for all area fractions considered. Thus, by means
of heterogeneous aggregation, it is possible to obtain a wide variety of
structures while more close-packed structures are formed for monodisperse 
colloidal aggregation.

\section*{Acknowledgments}
This work was supported by the Research Grants Council of the Hong Kong 
SAR Government under grant CUHK 4245/01P.
K.W.Y. acknowledges the hospitality received during his participation in 
the Adriatico Research Conference on Interaction and Assembly of 
Biomolecules, hosted by the International Center for Theoretical Physics 
at Italy, where the present work was initiated.
We acknowledge useful discussion with Dr. Jones T. K. Wan.

\begin{figure}[h]
\caption{Plot the time series of $g_0$ of aggregation with four panels (a)--(d), 
from initial to final time, both for one type and two types of particles 
(shown on the same figure).}
\end{figure}

\begin{figure}[h]
\caption{Plot the partial RDF's ($g_{AA}$, $g_{BB}$, $g_{AB}$) as well as the total 
RDF ($g_0$) with six panels (a)--(f), corresponding to the six area 
fractions studied, for the binary colloidal case.}
\end{figure}

\begin{figure}[h]
\caption{Plot the RDF ($g_0$)with six panels (a)--(f), 
corresponding to the six area fractions studied, 
for the monodisperse case.}
\end{figure}

\begin{figure}[h]
\caption{Plot the average coordination number $\langle Z\rangle$ versus area 
fraction both for one type and two types of particles.}
\end{figure}

\begin{figure}[h]
\caption{Plot the partial RDF's for the $u_1=20$ case. 
The $u_1=6$ case is plotted on the same figure for comparison.}
\end{figure}

\newpage
\centerline{\epsfig{file=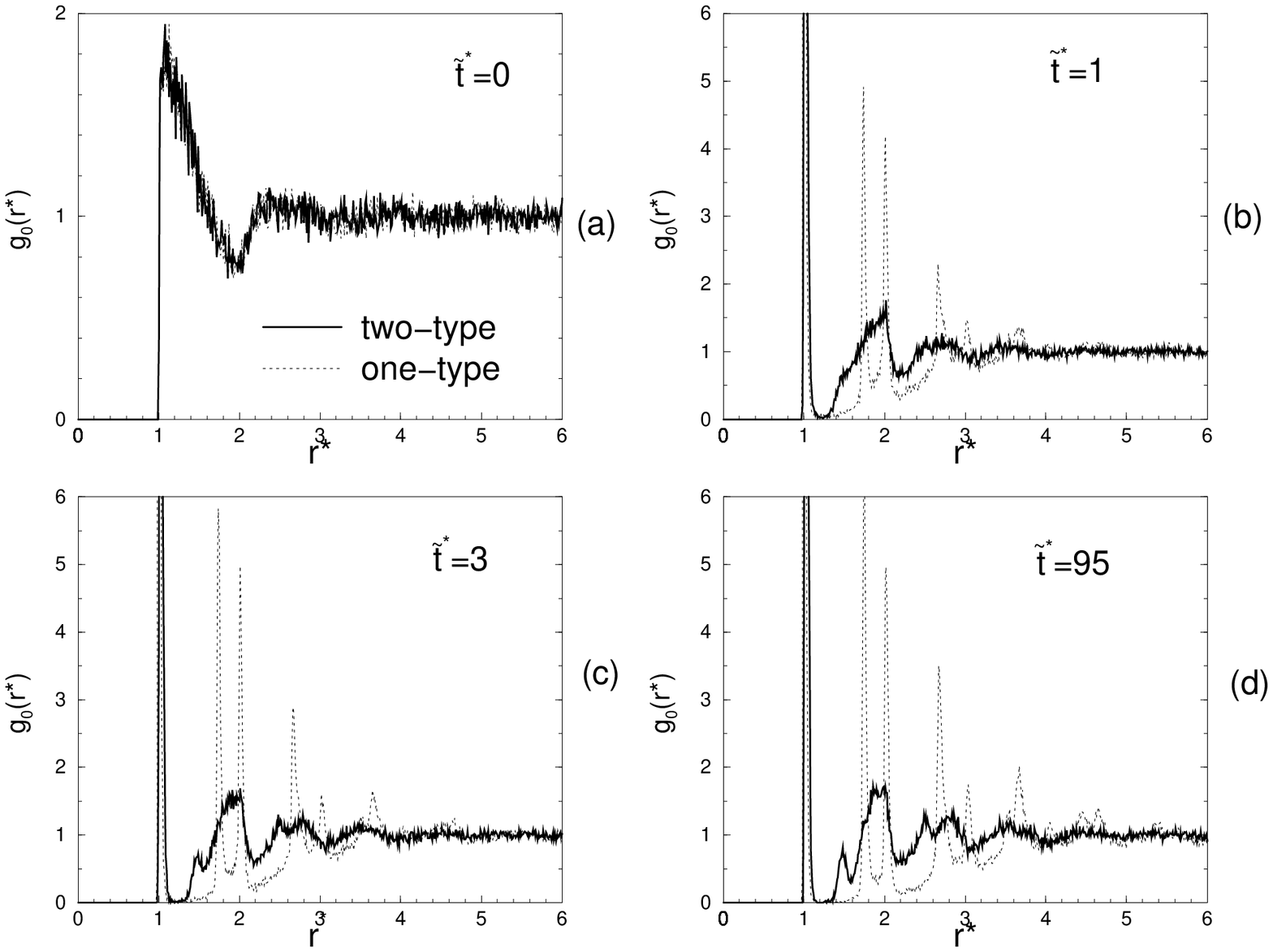,width=\linewidth}}
\centerline{Fig.1/Wong and Yu}

\newpage
\centerline{\epsfig{file=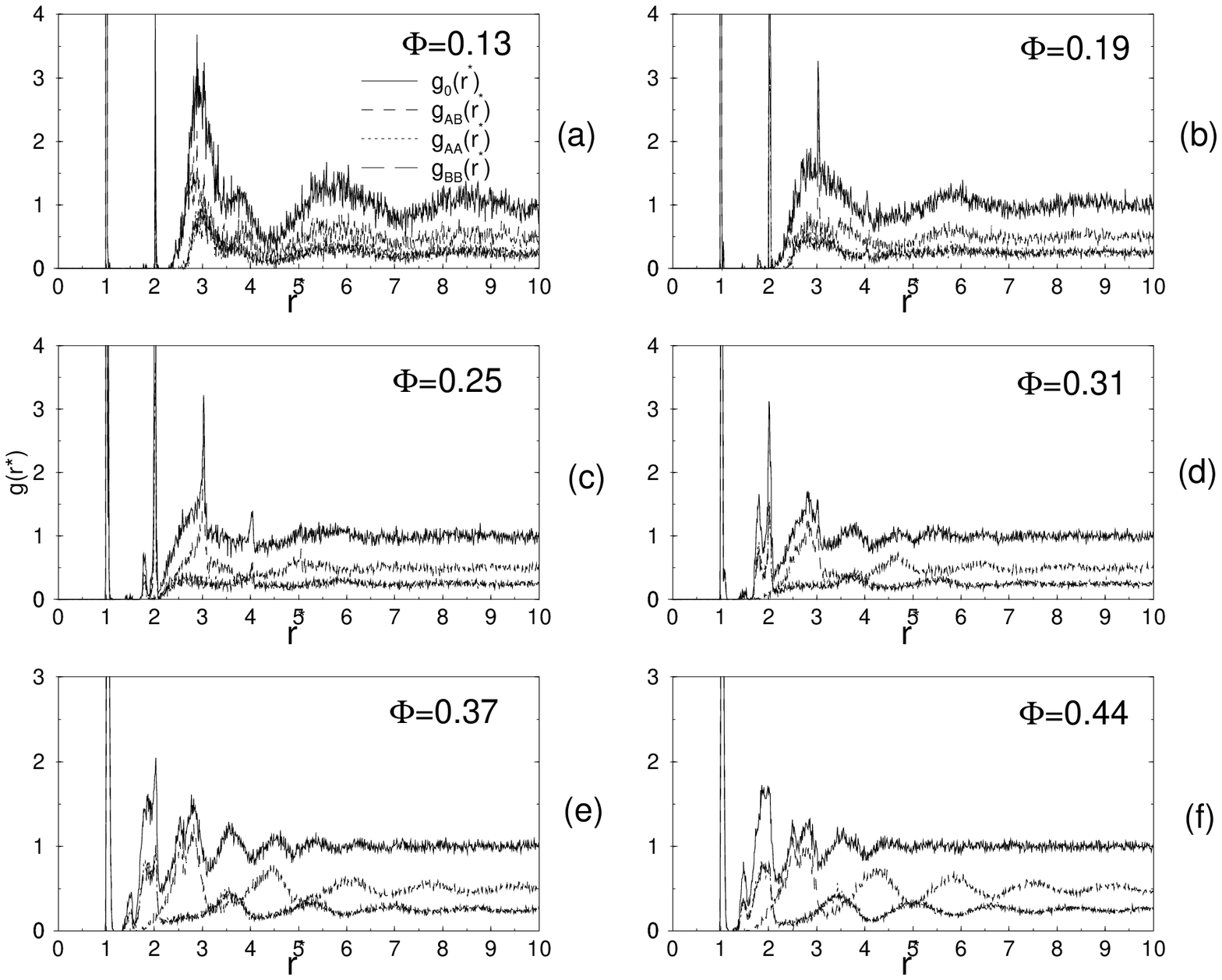,width=\linewidth}}
\centerline{Fig.2/Wong and Yu}

\newpage
\centerline{\epsfig{file=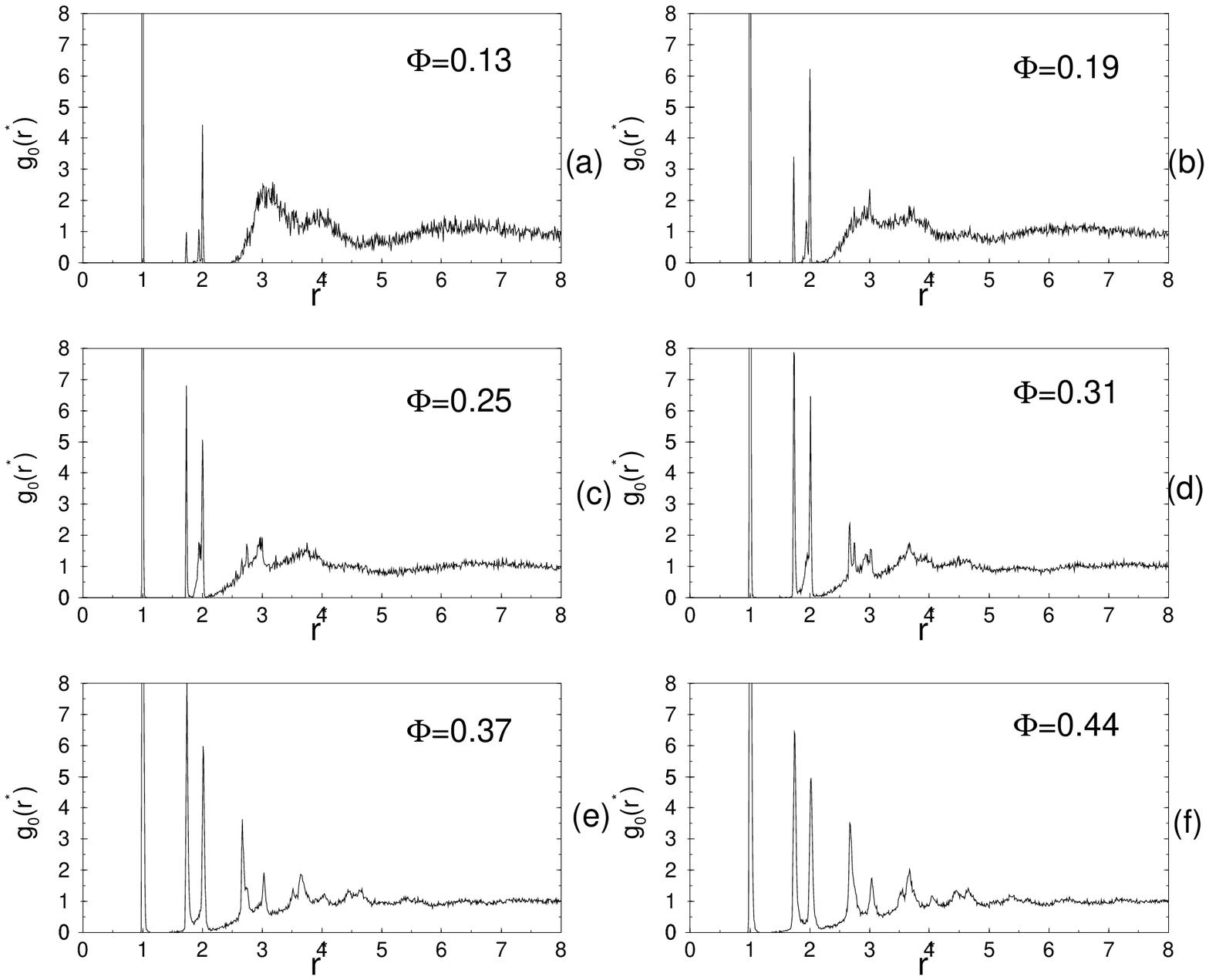,width=\linewidth}}
\centerline{Fig.3/Wong and Yu}

\newpage
\centerline{\epsfig{file=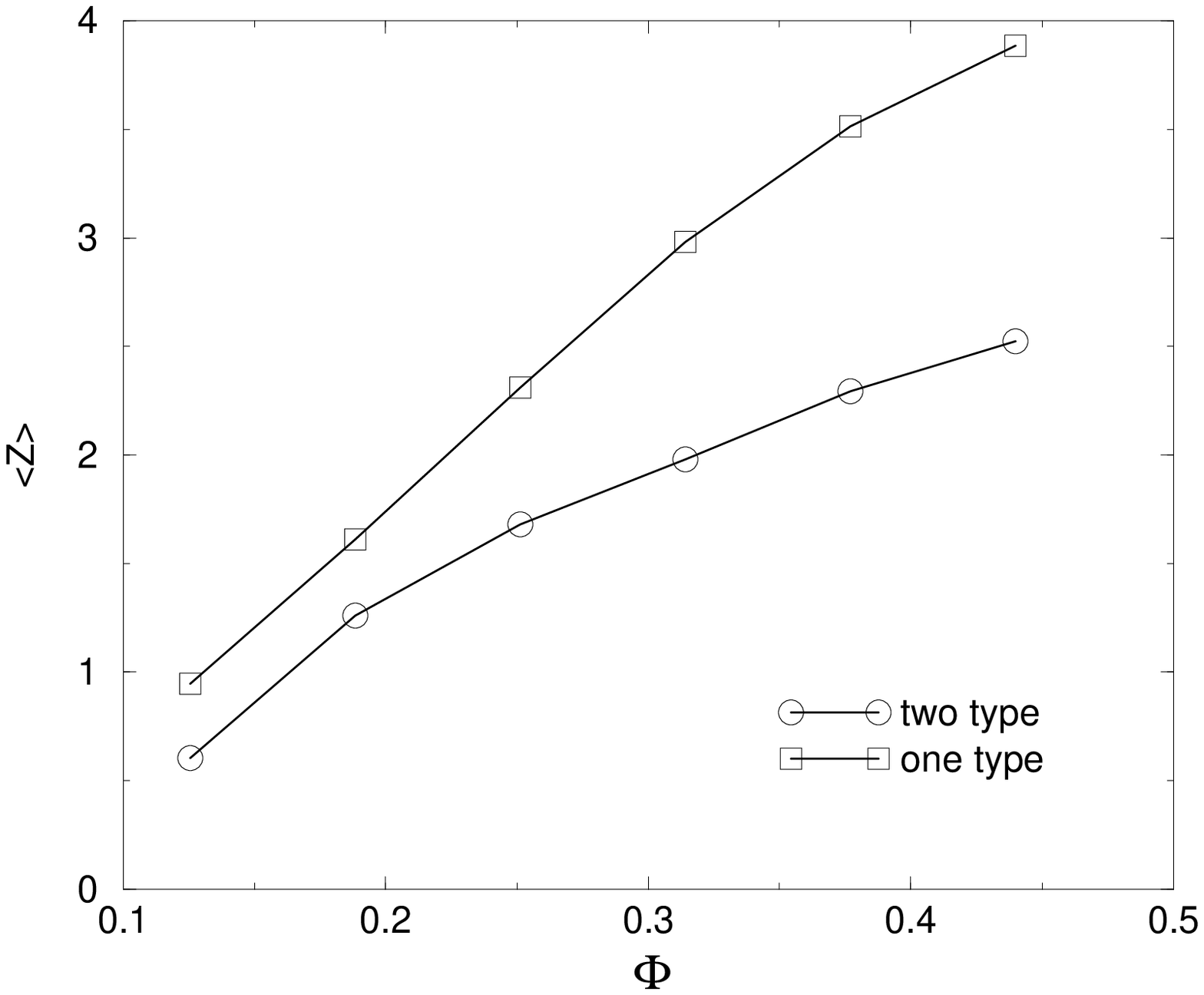,width=\linewidth}}
\centerline{Fig.4/Wong and Yu}

\newpage
\centerline{\epsfig{file=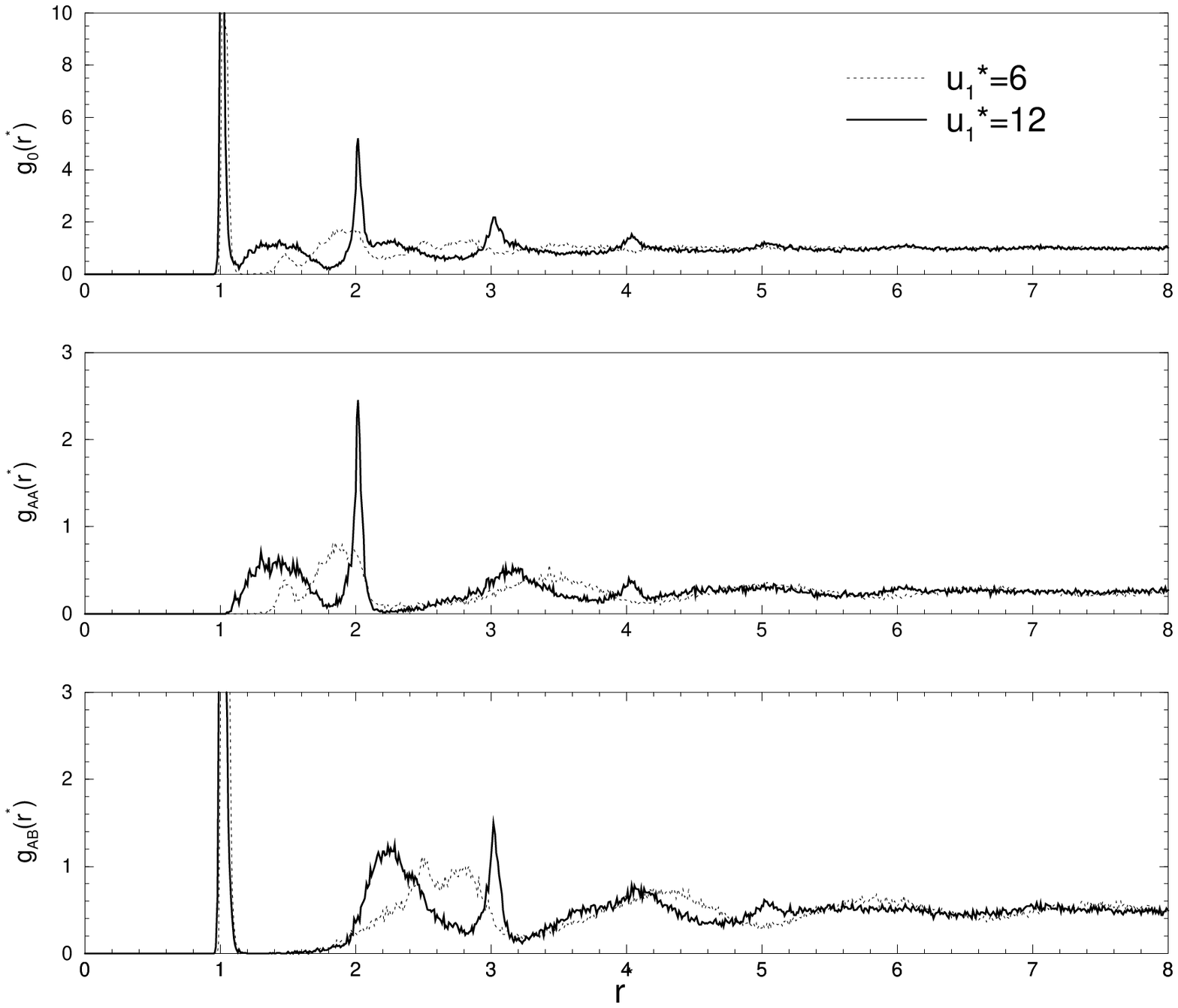,width=\linewidth}}
\centerline{Fig.5/Wong and Yu}
\end{document}